\documentclass[aps, prd, twocolumn, nofootinbib,floatfix]{revtex4}
\usepackage{subfigure}
\usepackage{graphicx}
\usepackage{dcolumn}
\usepackage{epsfig} 
\usepackage{amsmath}
\usepackage{amsfonts}
\usepackage{graphicx}
\usepackage{amssymb}
\begin{document}

\title{Small scale clustering of late forming dark matter}
\author{S. Agarwal$^{1}$, P.-S. Corasaniti$^{1}$, S. Das$^{2}$, Y. Rasera$^{1}$}
\affiliation{$^{1}$Laboratoire Univers et Th\'eories (LUTh), UMR 8102, CNRS, Observatoire de Paris, Universit\'e Paris Diderot, 5 Place Jules Janssen, 92190 Meudon, France}
\affiliation{$^2$Indian Institute of Astrophysics, 560034 Bangalore, India}

\date{\today}

\begin{abstract}
We perform a study of the nonlinear clustering of matter in the late-forming dark matter (LFDM) scenario in which dark matter results from the transition of a nonminimally coupled scalar field from radiation to collisionless matter. A distinct feature of this model is the presence of a damped oscillatory cutoff in the linear matter power spectrum at small scales. We use a suite of high-resolution N-body simulations to study the imprints of LFDM on the nonlinear matter power spectrum, the halo mass and velocity functions and the halo density profiles. The model largely satisfies high-redshift matter power spectrum constraints from Lyman-$\alpha$ forest measurements, while it predicts suppressed abundance of low-mass halos ($\sim 10^{9}-10^{10}$ h$^{-1}$ M$_\odot$) at all redshifts compared to a vanilla $\Lambda$CDM model. The analysis of the LFDM halo velocity function shows a better agreement than the $\Lambda$CDM prediction with the observed abundance of low-velocity galaxies in the local volume. Halos with mass $M\gtrsim 10^{11}$ h$^{-1}$ M$_\odot$ show minor departures of the density profiles from $\Lambda$CDM expectations, while smaller-mass halos are less dense, consistent with the fact that they form later than their $\Lambda$CDM counterparts.
\end{abstract} 

\maketitle

\section{Introduction}\label{intro}
The cold dark matter (CDM) scenario has been tremendously successful in reproducing observations of the distribution of matter on cosmic scales \cite{Efstathiou2002,Spergel2003,Tegmark2004,Clowe2006,PLANCK2011}. In spite of this remarkable success the origin of this invisible component is still not known. This is because cosmological observations shed no light on the particle physics nature of DM and only suggest that DM consists of an approximately pressureless component that clusters gravitationally since a few e-foldings before matter-radiation equality. 

The leading high-energy physics candidates to DM are weakly interacting massive particles (WIMPs). In principle these can be detected in underground laboratory experiments with a signal characterized by an annual modulation pattern. This has been thought to be a distinctive signature of DM direct detection against possible contamination from radioactive background noise. In fact the annual modulation is commonly considered a consequence of the orientation of the Earth's orbit along or against the flow of DM particles in the Galactic halo during the annual motion of the Earth around the Sun (for a review see Ref.~\cite{Freese2012}). A number of underground experiments such as DAMA/LIBRA \cite{DAMA2010}, CoGeNT \cite{CoGeNT2011} and CRESSST-II \cite{CRESSTII} have claimed the detection of DM particles with an annual modulation pattern. In contrast, the CDMS-II \cite{CDMSII2011}, XENON10 \cite{XENON2011} and more recently LUX \cite{LUX2013} collaborations have reported none. Searches for electroweak WIMPs at the Large Hadron Collider have also given negative results so far. More puzzling is the fact that the detections from the underground experiments constrain different regions of the WIMPs' parameter space \cite{Hooper2013}. Thus, it is possible that such contrasting outcomes may be hiding a much richer physics in the weak scale dark sector than previously thought or indicate the need of a completely new dark matter paradigm beyond weak scale WIMPs.

The small-scale clustering of matter in the Universe has emerged as the new arena to test the nature of DM. The discovery of a number of anomalies at small scales has cast doubts on the validity of the CDM hypothesis. The core-vs-cusp problem \cite{Moore1994,Kuzio2011} and the missing satellite problem (see e.g. \cite{Klypin1999,Moore1999}) have motivated the study of alternative scenarios beyond the CDM paradigm. Moreover, recent studies of the dynamical properties of the most luminous Milky Way dwarf spheroidal satellite galaxies have pointed to a new anomaly, the so-called too-big-to-fail problem \cite{BoylanKolchin2011,BoylanKolchin2012}. Baryonic processes in galaxy formation have been invoked as the natural solution to these discrepancies (see e.g. Refs.~\cite{Bullock2000,Benson2002,Somerville2002,Ricotti2002,Ricotti2005,Read2006,Libeskind2007,Maccio2010,Font2011,Pontzen2012,Teyssier2013}). As an example, baryon feedback and observational incompleteness can account for the missing satellite problem (see e.g. Refs.~\cite{Koposov2008,Guo2011}). Similarly, statistical variations in the predictions of subhalo abundances from N-body simulations combined with the uncertain value of the Milky Way virial mass may considerably alleviate the too-big-to-fail problem \cite{Wang2012,Purcell2012}. However, it is still unclear whether baryonic feedback models can provide a unique self-consistent explanation to the entirety of DM anomalies (see e.g. Refs.~\cite{Panarrubia2012,Ferrero2012,Garrison2013}). Furthermore, analyses of dwarf galaxies in the local field have shown that their abundance and properties differ significantly from the $\Lambda$CDM predictions \cite{Zavala2009,Papastergis2011,Kirby2014,Klypin2014,Garrison2014,Papastergis2015}. This is more difficult to reconcile with the CDM paradigm in terms of baryon feedback models given the fact that such isolated systems, contrary to satellites, undergo less complex processes. Therefore, it cannot be {\it a priori} excluded that such anomalies in the small-scale clustering of matter are (also) related to the unknown nature of DM (for a review see Ref.~\cite{Weinberg2013}) or a hint of broken scale invariance of the inflationary power spectrum at small scales \cite{KamionkowskiLiddle2000,Zentner2002,Zentner2003}.

Among the scenarios alternative to CDM, the existence of a warm dark matter (WDM) component has been motivated by particle physics models of sterile neutrinos (see e.g. Refs.~\cite{Dodelson1994,Fuller2003,Abazajian2006,Boyarsky2009a,Boyarsky2009b}). Differently from the CDM, WDM particles with mass of a few keV free-stream on a scale $\lesssim 100$ kpc due to velocity dispersion \cite{Zentner2003,Boyanovsky2008}. This causes a characteristic suppression of the linear matter density power spectrum at small scales and alters the properties of DM halos. Comparison with observations thus provides bounds on the DM particle mass. As an example, it was pointed out in Ref.~\cite{Maccio2012} that the solution to the core-vs-cusp problem requires WDM particles with mass $m_{\rm WDM}\sim 0.1$ keV. The authors of Refs.~\cite{Lovell2012,Lovell2014} have shown that WDM particles with mass in the range $1.5\lesssim m_{\rm WDM}[{\rm keV}]\lesssim 2$ can solve the too-big-to-fail problem. This also happens to be the optimal range for realistic sterile neutrino models \cite{Abazajian2014}. The effective thermal WDM mass range is severely constrained by the determination of the high-redshift matter power spectrum inferred from Lyman-$\alpha$ forest measurements \cite{Viel2005,Seljak2006,Viel2008}. For instance, the recent analysis by Viel {\sl et al.} \cite{Viel2013} indicates a lower limit $m_{\rm WDM}\gtrsim 3.3$ keV at a $2\sigma$ confidence level. Even assuming a particle mass within such a bound (e.g. $m_{\rm WDM}= 4$ keV) the authors of Ref.~\cite{Schneider2014} have shown that the dynamical properties of galactic WDM subhalos are indistinguishable from those of CDM, thus leaving the too-big-to-fail problem unsolved. Indeed, if the small-scale anomalies are entirely due to the particle physics nature of dark matter, then requiring their simultaneous solution poses strong constraints on DM models (see e.g. Ref.~\cite{Anderhalden2013}).

Another class of models alternative to the standard CDM hypothesis has developed around the possibility that DM particles have significant self-interactions \cite{Spergel2000}. This scenario, also known as self-interacting dark matter (SIDM), arises in the context of particle physics models of the DM sector (see e.g. Refs.~\cite{Feng2009,Arkani2009,Cline2014}). A key prediction of these models is the fact that DM particles have velocity-dependent self-scattering cross sections. These self-interactions, which can be assimilated to gravitational scalar forces, alter the nonlinear gravitational collapse, leaving characteristic signatures in the small-scale clustering of matter. N-body simulations of SIDM models have been presented in numerous studies. Recent analysis (see e.g. \cite{Vogelsberger2012,Zavala2013,Vogelsberger2014}) has shown that in a narrow range of the SIDM model parameter space the dynamics of subhalos in parent halos with mass about that of the Milky Way reproduces that of the dwarf-spheroidal galaxies in the Milky Way. At the same time, the subhalos possess core profiles $\mathcal{O}(1\,\textrm{kpc})$, thus simultaneously solving the core-cusp problem and the too-big-to-fail problem. On the other hand, in SIDM models the subhalo mass function remains identical to that of CDM, thus leaving the missing satellite problem unsolved. It has been shown in Refs.~\cite{Kamada2013,Boehm2014,Buckley2014} that such a class of models can provide a successful solution if an additional interaction of the DM component with radiation is considered. 

A different family of models consists of scenarios in which DM is the result of a decay or a phase transition process. For instance, the authors of Ref.~\cite{SigurdsonKamionkowski2004} have explored the possibility that neutral dark matter particles are the result of the decay of charged particles coupled to the photon-baryon plasma before recombination, while Ref.~\cite{Annika2010} has proposed a scenario in which early dark matter particles decay into a less massive species and a massless noninteracting one. In the latter case the resulting species eventually disrupt the formation of low-mass halos as investigated in Refs.~\cite{MWang2012,MWang2014}, while still satisfying bounds from Lyman-$\alpha$ observations \cite{MWang2013}.

Here, we consider a scenario in which the DM is consequence of a phase transition in the dynamics of a scalar field. In particular, we focus on \textit{late-forming dark matter} (LFDM) \cite{Das2011}, inspired by particle physics models of neutrino dark energy \cite{Fardon2006} which aim to provide a unified description of dark matter and dark energy. In LFDM models, DM particles are the result of a phase transition in the equation of state of a scalar field from radiation ($w\sim 1/3$) to matter ($w\sim 0$) (see Ref.~\cite{Das2012} for a different realization of this scenario). A distinctive characteristic of LFDM models is the presence of a damped oscillatory tail at the small scales of the linear matter power spectrum. However, contrary to other nonstandard DM models, the range of scales where such a distinctive feature occurs is set by the epoch of the phase transition (rather than the value of the DM particle mass or the amplitude of the self-interaction cross section). The earlier the transition, the smaller the scale where the suppression of power occurs. In this regard, models of ultralight axion (ULA) dark matter can also be seen as a form of late-forming dark matter where the axion field transitions from a vacuum state ($w\sim -1$) to matter ($w\sim 0$) when the field mass becomes comparable to the Hubble scale (see e.g. Ref.~\cite{MarshFerreira2010}). ULA dark matter alters the linear cosmic structure formation; thus it is constrained by observations of the large-scale structures as recently shown in \cite{Hlozek2014}. This occurs in LFDM only if the transition to DM state occurs after recombination, otherwise LFDM exclusively affects the linear matter power spectrum at small scales, thus leaving potentially observable features in the nonlinear structure formation. Given the similarities between the ULA and LFDM models, it is reasonable to expect that in some viable region of the ULA parameter space, the two scenarios may share the same phenomenology of the matter clustering at small scales, which is yet to be studied. In this paper we present a first study of the nonlinear structure formation of LFDM and assess the viability of this scenario using a series of high-resolution N-body simulations.

The paper is organized as follows: In Sec.~\ref{lfdm_cosmo} we recall the main features of the LFDM model. In Sec.~\ref{simulation_code} we describe the numerical methods, the generation of initial conditions and N-body simulation characteristics. In Sec.~\ref{pow} we analyze the nonlinear matter power spectrum. In Sec.~\ref{hmf} we describe the imprints on the halo mass function. In Sec.~\ref{profiles} we present the results from the halo density profiles. In Sec.~\ref{mcvf} we probe the circular velocity distribution, and we conclude in Sec.~\ref{conclu}.

\section{Cosmology of Late-Forming Dark Matter}\label{lfdm_cosmo}

In the following, we will briefly review the main features of the LFDM scenario that are relevant to the study of cosmic structure formation. We refer the reader to the original paper by Das and Weiner \cite{Das2011} for a detailed description of the model.

\begin{figure*}
\begin{centering}
\subfigure{\includegraphics[scale=0.4]{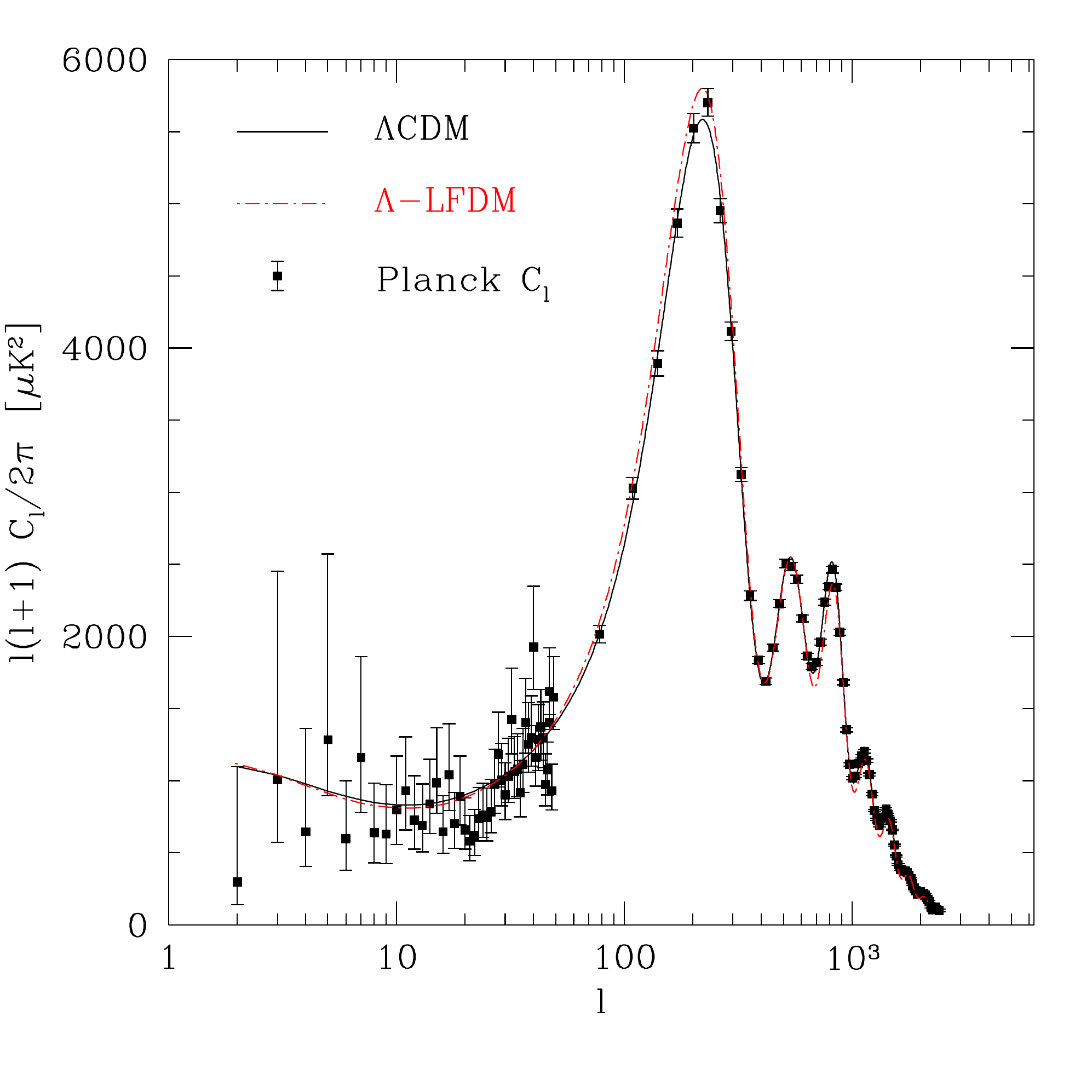}}\subfigure{\includegraphics[scale=0.4]{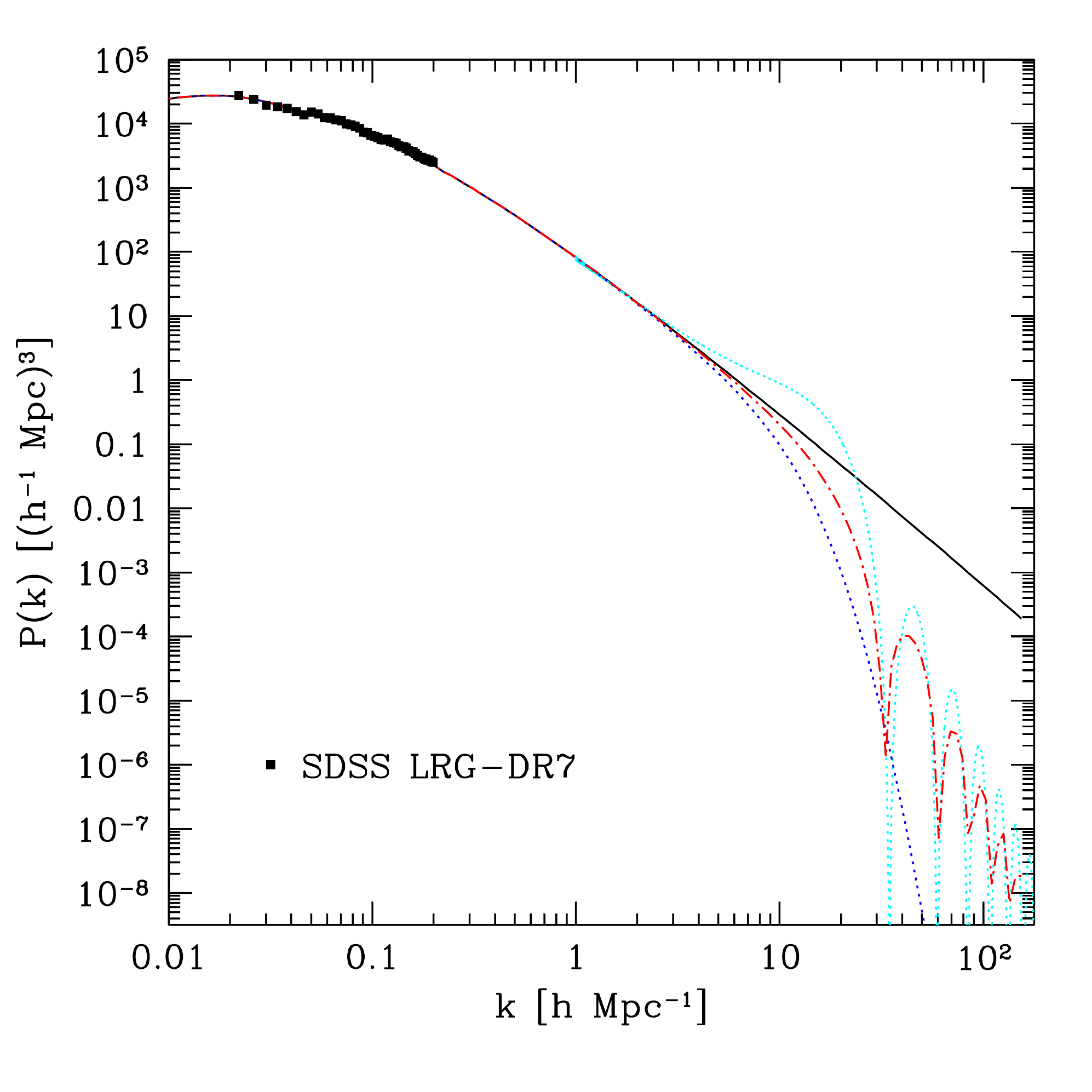}}
\caption{Left panel: CMB temperature anisotropy power spectrum for a flat $\Lambda$CDM model (black solid line) best fit to Planck data \cite{Planck2013cosmoconstr,Planck2013powerdata} (filled squares) and a flat $\Lambda$-LFDM model (red dot-dashed line) with identical cosmological parameters. Right panel: Linear matter density power spectra for the $\Lambda$CDM and $\Lambda$-LFDM models shown in the left panel. Data points correspond to LRG power spectrum from SDSS-DR7 \cite{Reid2010}. For illustrative purposes only we also show the linear power spectrum for a WDM model with thermal relic particle mass $m_{\rm WDM}=1.465$ keV (blue dotted line) and a broken-scale-invariant inflationary model (cyan dotted line).}\label{fig1}
\end{centering}
\end{figure*}

In this scenario, DM is the result of the dynamical evolution of a scalar field coupled to a thermal bath of relativistic particles (e.g. eV sterile neutrinos). At high temperature (early times) the field is trapped in a metastable state and behaves as a dark radiation fluid with an equation of state $w\sim 1/3$. As temperature drops due to the cosmic expansion, a new lower energy minimum appears and the field rolls into the lower energy state while oscillating around the true minimum of the scalar potential, thus behaving as a collisionless DM component ($w\sim 0$). Because of the coupling to the thermal bath of relativistic particles, deep in the radiation era (before phase transition) the scalar field density fluctuates through damped oscillations (similarly to the baryon acoustic oscillations in the photon-baryon plasma during the tight-coupling regime). After the transition, LFDM behaves as standard CDM, and the damped oscillations remain imprinted in the density fluctuation power spectrum. This damped oscillatory pattern is characteristic of other alternative DM scenarios in which DM is coupled to radiation (see e.g. Refs.~\cite{Boehm2002,Racine2013,Wilkinson2014}). However, in the LFDM model the range of scales carrying the imprint of damped oscillation depends on the redshift of the phase transition, $z_t$. The later the phase transition, the larger the scales where the power spectrum exhibits a damped oscillatory tail. The absence of these features in the large-scale power spectrum of galaxies imposes constraints on the redshift of the transition. In particular, a recent analysis indicates that $z_t>10^5$ \cite{Sarkar2014}. Since after the phase transition LFDM behaves as a collisionless component, the linear cosmic structure formation is indistinguishable from that of the standard $\Lambda$CDM model. However, in order to produce sufficient dark matter abundance, an excess of scalar dark radiation (in addition to that of neutrinos and photons) is needed at early times. As shown in Ref.~\cite{Sarkar2014}, a small excess contributing to $\Delta N_{\rm eff}\sim 0.01-0.1$, thus well within the CMB constraints \cite{Planck2013cosmoconstr}, is sufficient to give the right amount of dark matter particles.

As an example, in the left panel of Fig.~\ref{fig1} we plot the cosmic microwave background (CMB) temperature anisotropy power spectrum for a flat $\Lambda$-LFDM model with cosmological constant against a $\Lambda$CDM model best fit to Planck data \cite{Planck2013cosmoconstr,Planck2013powerdata}. In this specific LFDM model the phase transition occurs at $z_t=1.5\times 10^{6}$, and the cosmological parameters are set to the values of the standard $\Lambda$CDM model. We have computed the CMB and matter power spectrum of the LFDM model using a modified version of the {\sc camb} code \cite{Lewis2000} to solve the dynamics of the coupled linear perturbation equations associated with the LFDM component. We may notice that even without a model parameter optimization requiring a Monte Carlo likelihood analysis of available data, the choice of the $\Lambda$-LFDM model parameters provides a good fit to the Planck measurements (the small differences arise essentially from the small excess of dark radiation density before matter-radiation equality). In the right panel of Fig.~\ref{fig1}, we plot the corresponding linear matter power spectra at redshift $z=0$ against the luminous red galaxy (LRG) power spectrum from the Sloan Digital Sky Survey seventh data release (SDSS DR7) \cite{Reid2010}. As we can see the spectra are identical on the large scales, while differences are only present at $k\gtrsim 10$ h Mpc$^{-1}$ due to LFDM damped oscillations on nonlinear scales. The earlier the phase transition (i.e. higher $z_t$), the smaller the power spectrum cutoff scale, and the damped part of the spectrum shifts towards larger wave numbers.

It is worth mentioning that the linear LFDM power spectrum considered here resembles that of the models studied in Refs.~\cite{Boehm2014,Buckley2014}. Nonetheless, there are a few noticeable differences. In Ref.~\cite{Boehm2014} the authors have considered nonstandard models with DM-photon interactions characterized by a linear power spectrum that in the range $10<k[\,{\rm h\, Mpc^{-1}}]<100$ has a damped oscillatory pattern with a slope different from that of the LFDM case (see Fig.~1 in Ref.~\cite{Boehm2014}), while in Ref.~\cite{Buckley2014} the amplitude of the damped oscillations is significantly more suppressed than in our model realization (see Fig.~1 in Ref.~\cite{Buckley2014}). Certain realizations of mixed cold and warm dark matter models such as those considered in Refs.~\cite{Boyarsky2009b,Maccio2013} also have spectra that approximate the envelope of the LFDM spectrum; however, they have different power distribution in the range $10<k[\,{\rm h\, Mpc^{-1}}]<100$. Indeed, the slope of the suppressed part of the spectrum is a key aspect that differentiates the nonlinear structure formation of scenarios whose phenomenology depends solely on the presence of a cutoff in the linear matter power spectrum. For instance, in the right panel of Fig.~\ref{fig1} we plot the linear spectra of a WDM thermal relic particle with mass $m_{\rm WDM}=1.465$ keV characterized by a cutoff scale similar to that of LFDM, and of a broken-scale-invariance (BSI) inflationary model \cite{Starobinsky1992,Lesgourgues1998} as the one considered in Ref.~\cite{KamionkowskiLiddle2000} with highly tuned parameters\footnote{The primordial spectrum for BSI models \cite{Starobinsky1992,Lesgourgues1998} has a universal form and depends only on two parameters, the cutoff scale and the amplitude of the power suppression $p$. In Refs.~\cite{Zentner2002,Zentner2003} it was assumed $p=4$, while to closely reproduce the LFDM spectrum we set $p=100$.} such as to have the same cutoff scale and a slope of power suppression similar to that of the LFDM model. We can see that the spectra still carry differences that are likely to alter the nonlinear structure formation. In particular, in the case of the BSI spectrum, the model predicts an excess of power compared to the $\Lambda$CDM case before the cutoff; thus it is likely to be highly constrained by high-redshift Lyman-$\alpha$ measurements. This is not the case for the WDM model, which differs from the LFDM beyond the cutoff scale, where the power exponentially drops several orders of magnitude below the envelope of LFDM damped oscillations. Even considering a slightly larger thermal mass particle, this will only shift the exponential tail of the WDM spectrum to higher wave numbers, while still remaining more suppressed than the LFDM prediction due to the shallower slope of the LFDM envelope. Hence, it is reasonable to expect that such differences may manifest with different predictions of the abundance and properties of low-mass halos. It is for these very reasons that, despite the numerous numerical studies of the nonlinear clustering of WDM models, it is opportune to investigate the phenomenology of the LFDM scenario and the like.

\section{N-body Simulations}\label{simulation_code}
We run a series of high-resolution N-body simulations using {\sc ramses} \cite{Teyssier2002}, an adaptive mesh refinement (AMR) code with a tree-based data structure in which particles are evolved using a particle-mesh solver and the Poisson equation is solved using a multigrid method \cite{Guillet2011}. 

We generate the initial conditions with the code {\sc mpgrafic} \cite{Prunet2008}, which uses the Zel'dovich approximation. We set the initial redshift such that the standard deviation of the initial density field smoothed on the scale of the coarse grid $\Delta_x^{\rm coarse}$ is $\sigma(\Delta_x^{\rm coarse})=0.02$. With this choice, the initial redshift of the simulations is sufficiently large to suppress spurious effects due to transients \cite{Crocce2006}. 

We assume a flat $\Lambda$-LFDM model with phase transition redshift $z_t=1.5\times 10^6$ and characterized by the linear power spectrum shown in Fig.~\ref{fig1}. The cosmological parameters are set to the following values: $\Omega_m=0.3$, h $=0.7$, $\sigma_8=0.8$, $n_s=0.96$ and $\Omega_b=0.046$. In addition, we also consider a flat $\Lambda$CDM model with identical parameters.   

The simulations of the $\Lambda$-LFDM model consist of a box of $(27.5\,{\rm h^{-1} Mpc})^3$ volume with $512^3$ particles, a similar volume with $1024^3$ particles, an intermediate box of $(64\,{\rm h^{-1} Mpc})^3$ with $512^3$ particles and a large box of $(110\,{\rm h^{-1} Mpc})^3$ volume with $2048^3$ particles. This simulation suite enables us to control numerical systematic errors due to mass resolution and volume effects. The characteristics of the simulations are summarized in Table~\ref{table_sim}. We can see that for the highest mass resolution runs the spatial resolution of the simulations at the level of the coarse grid varies from $27\,{\rm h^{-1} kpc}$ to $54\,{\rm h^{-1} kpc}$. However, the actual resolution of the simulations is much higher due to the AMR scheme used in the {\sc ramses} code. For instance, in the case of the $(27.5\,{\rm h^{-1} Mpc})^3$ volume simulation with $1024^3$ particles we have up to six refinement levels of the coarse grid which are triggered during the run; thus the densest objects are resolved with a spatial resolution of $\approx 3.4\,{\rm h^{-1} kpc}$. 

In addition to the LFDM model simulations, we have run a $(27.5\,{\rm h^{-1} Mpc})^3$ volume with $512^3$ particles and one with $1024^3$ particles for a flat $\Lambda$CDM model with identical cosmological parameters (with $z_{\rm ini}=424$ and $z_{\rm ini}=502$, respectively) and the same phase of the initial conditions, which we use to evaluate differences with respect to the small-scale clustering of $\Lambda$-LFDM.

The simulations were run on the ADA supercomputer of the Institute for Development and Resources in Intensive Scientific Computing (IDRIS). In particular, the LFDM simulation of $(110\,{\rm h^{-1} Mpc})^3$ volume with $2048^3$ particles was run on 3000 Intel Sandy Bridge E5-4650 processors for a total running time of $3\times 10^6$ hours.

\begin{table}[t]
\begin{tabular}{ccccc}
\hline
L (h$^{-1}$ Mpc) & $N_p$ & $z_{\rm ini}$ &$m_p$ (h$^{-1}$ M$_\odot$) & $\Delta_x^{\rm coarse}$ (h$^{-1}$ kpc) \\
\hline
$27.5$ & $512^3$ & 320 & $1.3\times 10^7$ & 54  \\
$27.5$ & $1024^3$ & 324 & $1.6\times 10^6$ & 27  \\
$64$ & $512^3$ & 303 & $1.6\times 10^8$ & 125\\
$110$ & $2048^3$ & 320 & $1.3\times 10^7$ & 54 \\
\hline
\end{tabular}
\caption{$\Lambda$-LFDM N-body simulation characteristics. L is the simulation box length, $N_p$ is the number of N-body particles, $z_{\rm ini}$ is the initial redshift of the simulation, $m_p$ the mass resolution, and $\Delta_x^{\rm coarse}$ is the spatial resolution of the coarse grid.}\label{table_sim}
\end{table}

\begin{figure*}
\begin{centering}
\subfigure{\includegraphics[scale=0.4]{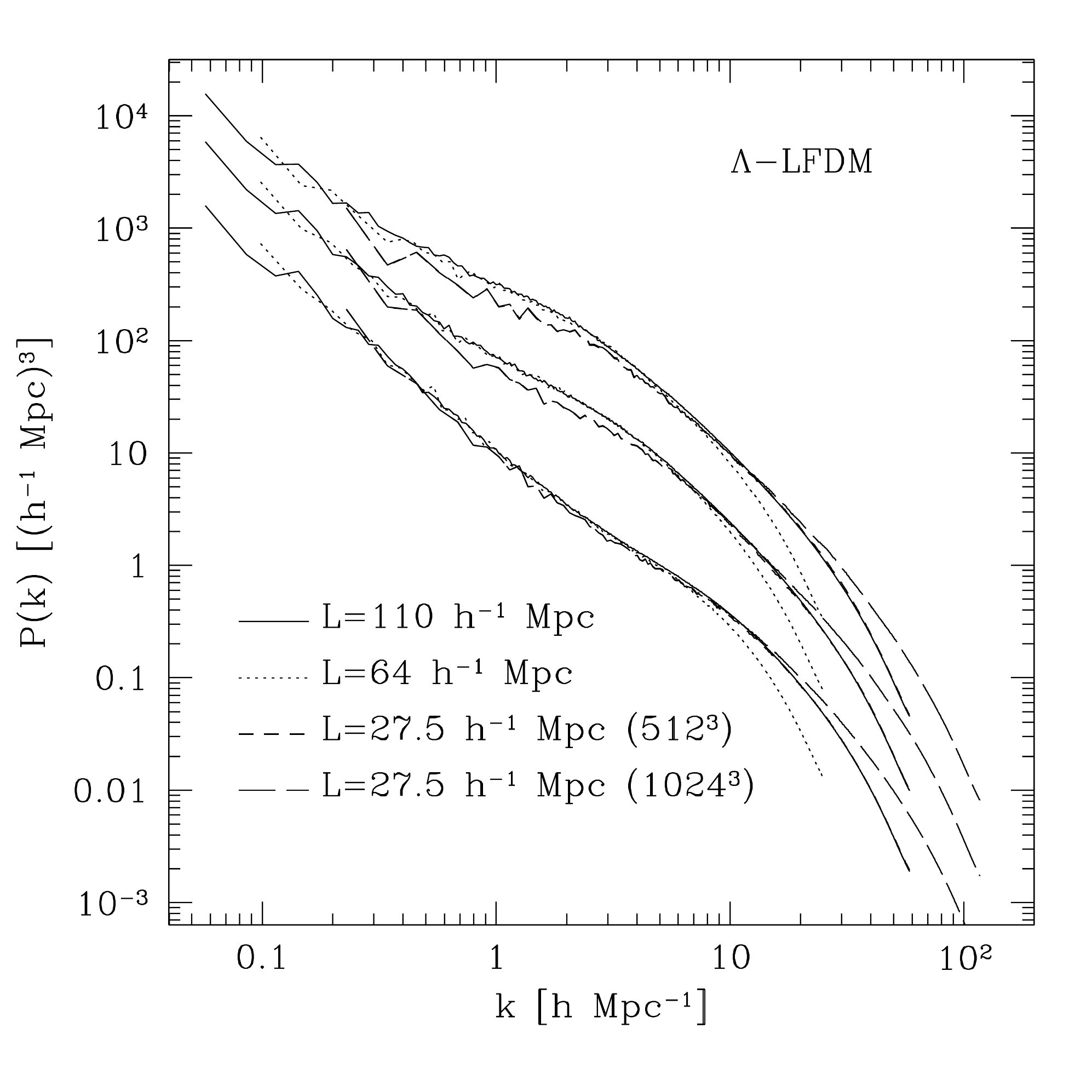}}\subfigure{\includegraphics[scale=0.4]{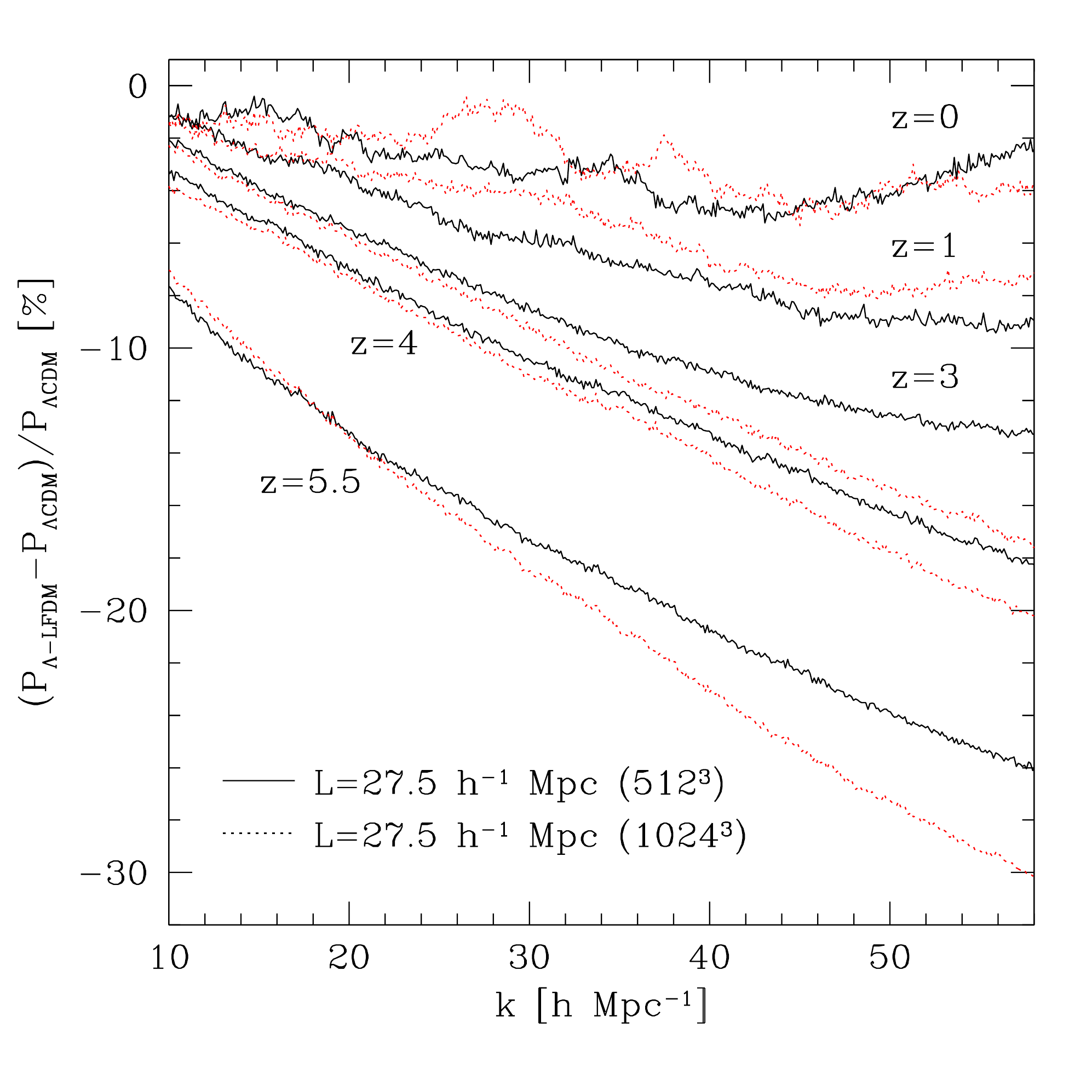}}
\caption{Left panel: Nonlinear matter power spectrum of the $\Lambda$-LFDM model from the simulations of box length $L=110\,{\rm h^{-1} Mpc}$ with $2048^3$ particles (solid line), $L=64\,{\rm h^{-1} Mpc}$ with $512^3$ particles (dotted line), and $L=27.5\,{\rm h^{-1} Mpc}$ with $512^3$ (short dashed line) and $1024^3$ (long dashed line) particles at $z=0,1$ and $3$ (top to bottom). Right panel: Relative difference of the nonlinear matter power spectrum of the $\Lambda$-LFDM model with respect to the $\Lambda$CDM case at $z=0,1,3,4$ and $5.5$ (top to bottom) from the simulations of $(27.5\,{\rm h^{-1} Mpc})^3$ volume with $512^3$ (black solid line) and $1024^3$ (red dotted line) particles.}\label{fig2}
\end{centering}
\end{figure*}

\section{NonLinear Matter Power Spectrum}\label{pow}
We compute the matter power spectrum with the code {\sc powergrid} \cite{Prunet2008}. This computes the power spectrum by performing a Fourier transform of the density field in band powers $\Delta k=2\pi/L$, where $L$ is the simulation box length. We correct for the smoothing effect due to the cloud-in-cell (CIC) algorithm used to estimate the density field from the particle distribution. To be conservative, we restrict the scope to wave numbers below the Nyquist frequency of the coarse grid.

In the left panel of Fig.~\ref{fig2}, we plot the nonlinear power spectra at $z=0,1$ and $3$ from the four $\Lambda$-LFDM simulations which differ for volume size and mass resolution. For clarity, it is convenient to focus on the $z=0$ case, with similar trend at higher redshifts. First, we may notice that the spectra of the large (solid line) and intermediate (dotted line) volumes are fairly in agreement at $k<6$ h Mpc$^{-1}$, while in the same range the amplitude of the spectrum of the smaller simulation box (dashed line) is slightly lower. This is due to finite volume effects as well as the choice of the initial phase (see e.g. Refs.~\cite{Heitmann2010,Rasera2014,Orban2014}). Volume effects are negligible at large $k$; in particular for $k > 13$ h Mpc$^{-1}$, the spectrum of the $(27.5\,{\rm h^{-1} Mpc})^3$ volume simulation with $512^3$ particles and that of the $(110\,{\rm h^{-1} Mpc})^3$ box differ by less than $2\%$. On the other hand, at larger wave numbers (near the Nyquist frequency of the simulations), we can see the systematic suppression of power due to mass resolution errors for the lower resolution runs compared to the higher one. At small scales, this is the dominant source of numerical uncertainty; in particular, in the range $10<k[{\rm h\,Mpc^{-1}}]<50$, the spectra of the $(27.5\,{\rm h^{-1} Mpc})^3$ volume simulation with $512^3$ and that with $1024^3$ particles differ by a few percent in the low-end limit and up to $50\%$ in the high end. However, since we are interested in the relative difference between the LFDM and $\Lambda$CDM spectra, we expect (and show in the right panel of Fig.~\ref{fig2}) the ratio to be less affected by mass resolution errors.

We plot in the right panel of Fig.~\ref{fig2} the relative difference of the nonlinear matter power spectrum of the $\Lambda$-LFDM model with respect to the $\Lambda$CDM case for $k\gtrsim 10$ h Mpc$^{-1}$ at $z=0,1,3,4$ and $5.5$ from the $(27.5\,{\rm h^{-1} Mpc})^3$ volume simulations. At $z=5.5$, the runs with $512^3$ and $1024^3$ particles differ from less than a percent at $k=10$ h Mpc$^{-1}$ and increase up to $4\%$ at the high end of the interval. These differences due to mass resolution are clearly subdominant compared to the relative differences between the models. Volume effects in this range of wave numbers are expected to be negligible. Hence, we are confident the differences seen between $\Lambda$-LFDM and $\Lambda$CDM are not due to numerical artifacts. Notice that any signature of the LFDM damped oscillations in the initial linear power spectrum has been completely erased.

Constraints on the high-redshift matter power spectrum ($3<z<5.4$) inferred from the recent Lyman-$\alpha$ measurements \cite{Viel2013} indicate that the nonlinear power spectrum for WDM thermal relic particles of mass $m_{\rm wdm}>2.5$ keV (at $3\sigma$ confidence level) may deviate from $\Lambda$CDM by no more than $\sim 5\%$ at $k=10$ h Mpc$^{-1}$ (as can be inferred from Fig.~1 in Ref.~\cite{Viel2013}). As shown in the right panel of Fig.~\ref{fig2}, the $\Lambda$-LFDM model considered here is consistent with these bounds for $z\lesssim 4$, while models with slightly higher $z_t$ will largely evade these constraints in the entire high-redshift range.

\section{Halo Mass Function}\label{hmf}
\subsection{Numerical convergence analysis}
We detect DM halos using the code pFoF \cite{Roy2014} based on the friend-of-friend algorithm \cite{Davis1985}, which identifies halos as group of particles characterized by an intraparticle distance smaller than a given linking length parameter $b$. In our analysis we set this to the standard value $b=0.2$ and consider only halos with more than $100$ particles unless specified otherwise. 

Spurious low-mass halos due to artificial fragmentation of the DM density field occur in models with a sharp cutoff in the initial power spectrum (see e.g. Refs.~\cite{Gotz2002,Gotz2003,Wang2007}). These alter the predictions of the low-mass end of the mass function and contribute as a dominant source of systematic errors in the mass range where differences among DM scenario are mostly relevant to observational tests. We have taken fragmentation into account in analysis of the $\Lambda$-LFDM model presented here and refer the reader to Ref.~\cite{Agarwal2015} for a dedicated study on the removal of spurious halos. Our spurious halo identification method relies on the fact that since spurious halos originate from the fragmentation of the density field, these are highly nonspherical groups of particles characterized by extreme values of the spin and shape parameters while exhibiting large deviation from the virial theorem. Thus, in the range of masses where mass resolution effects are subdominant, spurious halos can be removed by simply retaining those halos that approximately satisfy the virial condition and checking that the distribution of the spin and shape parameters of the remaining halos have well-behaved tails.  

We compute the mass function as
\begin{equation}
\frac{dn}{d\ln{M}}=\frac{1}{L^3}\frac{N}{\Delta\ln{M}},
\end{equation}
where $L$ is the simulation box length and $N$ is the number of halos in a mass bin of size $\Delta\ln{M}$. Throughout this paper, $\log$ and $\ln$ denote base-10 and base-$e$ logarithms, respectively.

\begin{figure}
\includegraphics[scale=0.4]{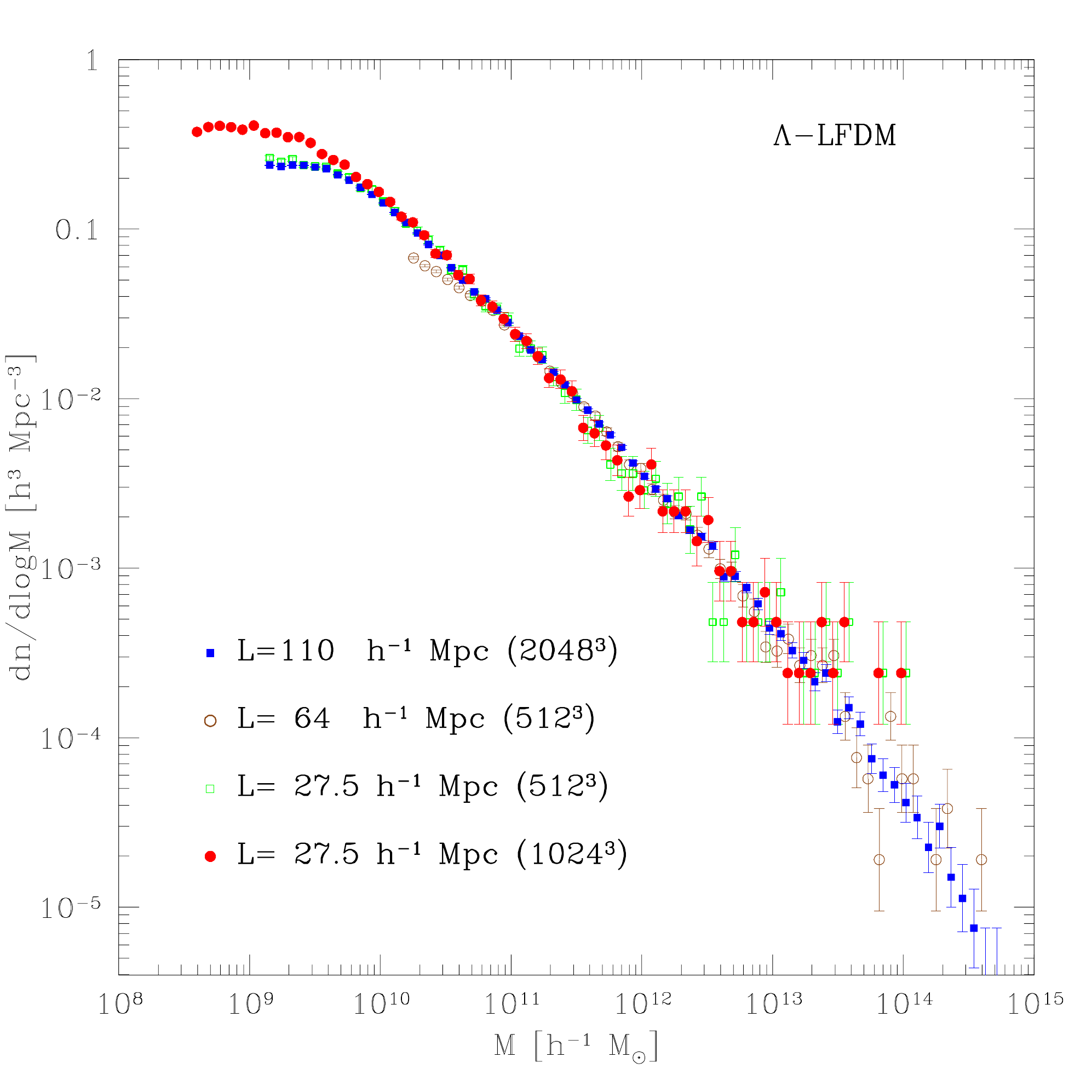}
\caption{Halo mass function of the $\Lambda$-LFDM model at $z=0$ in bins of size $\Delta{M}/M=0.1$ from simulations with volumes of $(110\,{\rm h^{-1} Mpc})^3$ with $2048^3$ particles (blue filled squares), $(64\,{\rm h^{-1} Mpc})^3$ with $512^3$ particles (brown open circles), and $(27.5\,{\rm h^{-1} Mpc})^3$ with $512^3$ (green open squares) and $1024^3$ (red filled circles) particles for halos with at least $100$ particles. Error bars are given by Poisson errors.}\label{fig3}
\end{figure}

In Fig.~\ref{fig3} we plot the mass function of the $\Lambda$-LFDM model at $z=0$ from simulations with volumes of $(110\,{\rm h^{-1} Mpc})^3$ with $2048^3$ particles (blue filled squares), $(64\,{\rm h^{-1} Mpc})^3$ with $512^3$ particles (brown open circles), $(27.5\,{\rm h^{-1} Mpc})^3$ with $512^3$ (green open squares) and $1024^3$ (red filled circles) particles in bins of size $\Delta{M}/M=0.1$. In all four cases we can see that the high-mass end of the mass function is characterized by a large level of scatter due to finite volume effects. Mass resolution effects are relevant in the low-mass end; this can be seen in the case of the intermediate simulation box. for which the mass function deviates at more than a $5\%$ level for $M<5\times 10^{10}$ h$^{-1}$ M$_\odot$ from the higher-resolution runs. Likewise, the comparison between the mass functions of the $(27.5\,{\rm h^{-1} Mpc})^3$ volume simulation with $512^3$ particles and that with $1024^3$ shows deviation at more than a $5\%$ level for $M<5\times 10^{9}$ h$^{-1}$ M$_\odot$. It is worth noticing the drop of the mass function from the $(27.5\,{\rm h^{-1} Mpc})^3$ volume run with $1024^3$ at low masses; this is consistent with the small-scale cutoff in the linear matter power spectrum. Overall, we find convergence at  the $5\%$ level in the mass range $5\times10^9\lesssim M\,[\rm h^{-1} M_\odot]\lesssim 10^{11}$ for the mass functions from the $(27.5\,{\rm h^{-1} Mpc})^3$ and $(110\,{\rm h^{-1} Mpc})^3$ volume simulations.

\begin{figure}
\includegraphics[scale=0.4]{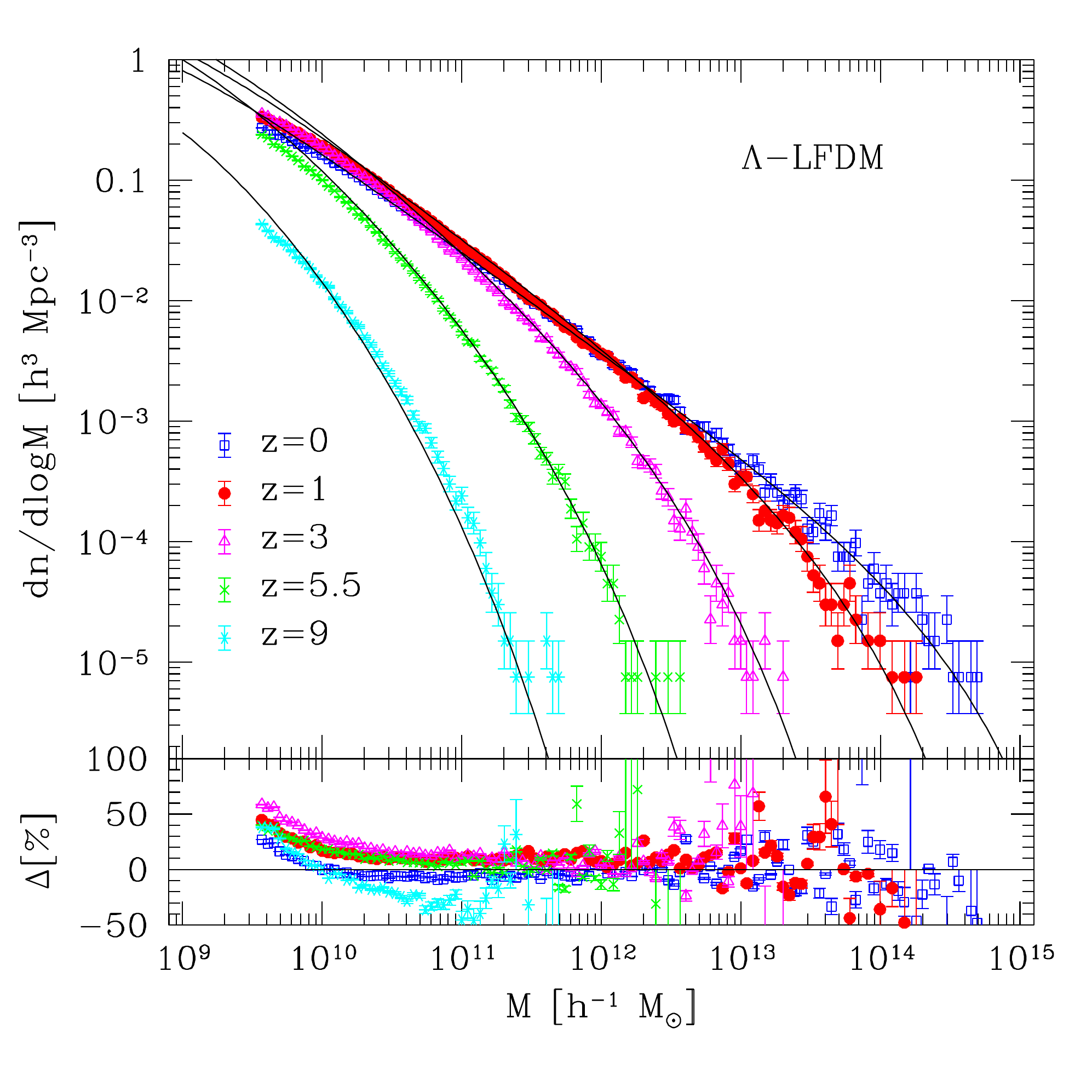}
\caption{Redshift evolution of the $\Lambda$-LFDM halo mass function at $z=0$ (blue open squares), $1$ (red solid circles), $3$ (magenta open triangles), $5.5$ (green stars) and $9$ (cyan asterisks) from the $(110\,{\rm h^{-1} Mpc})^3$ volume simulation. Error bars are given by Poisson errors. The solid lines show the mass function assuming the Sheth-Tormen multiplicity function with parameters calibrated at $z=0$.}\label{fig4}
\end{figure}

\subsection{Redshift evolution and cosmological imprints}
In Fig.~\ref{fig4} we plot the redshift evolution of the $\Lambda$-LFDM mass function from the $(110\,{\rm h^{-1} Mpc})^3$ volume simulation at $z=0,1,3,5.5$ and $9$. Here, to be conservative, we limit the mass range to halos with at least $350$ particles. For comparison, we also plot the Sheth-Tormen (ST) \cite{ST1999} mass function calibrated at $z=0$ and extrapolated to higher redshifts (see the Appendix). 

\begin{figure*}
\begin{centering}
\subfigure{\includegraphics[scale=0.4]{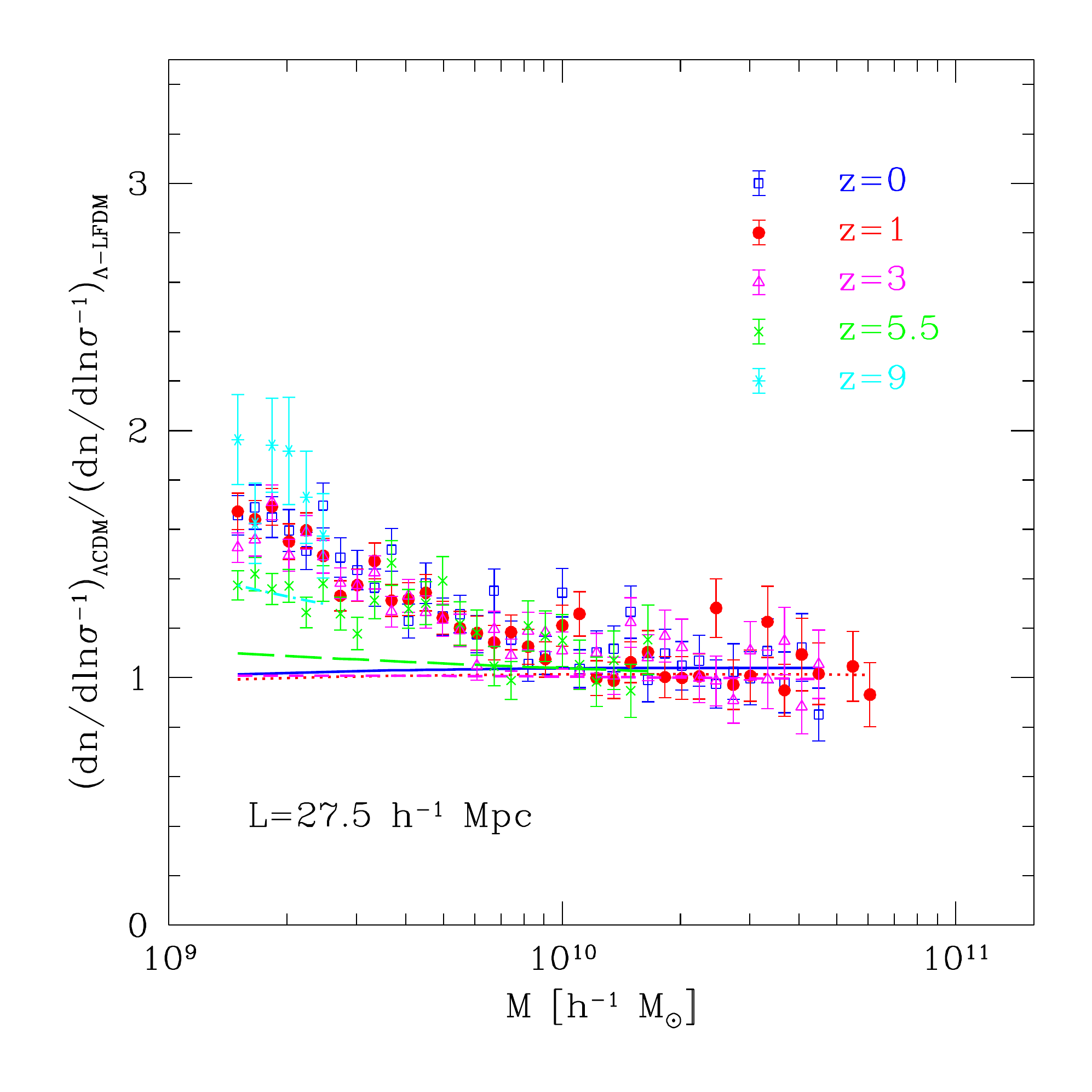}}\subfigure{\includegraphics[scale=0.4]{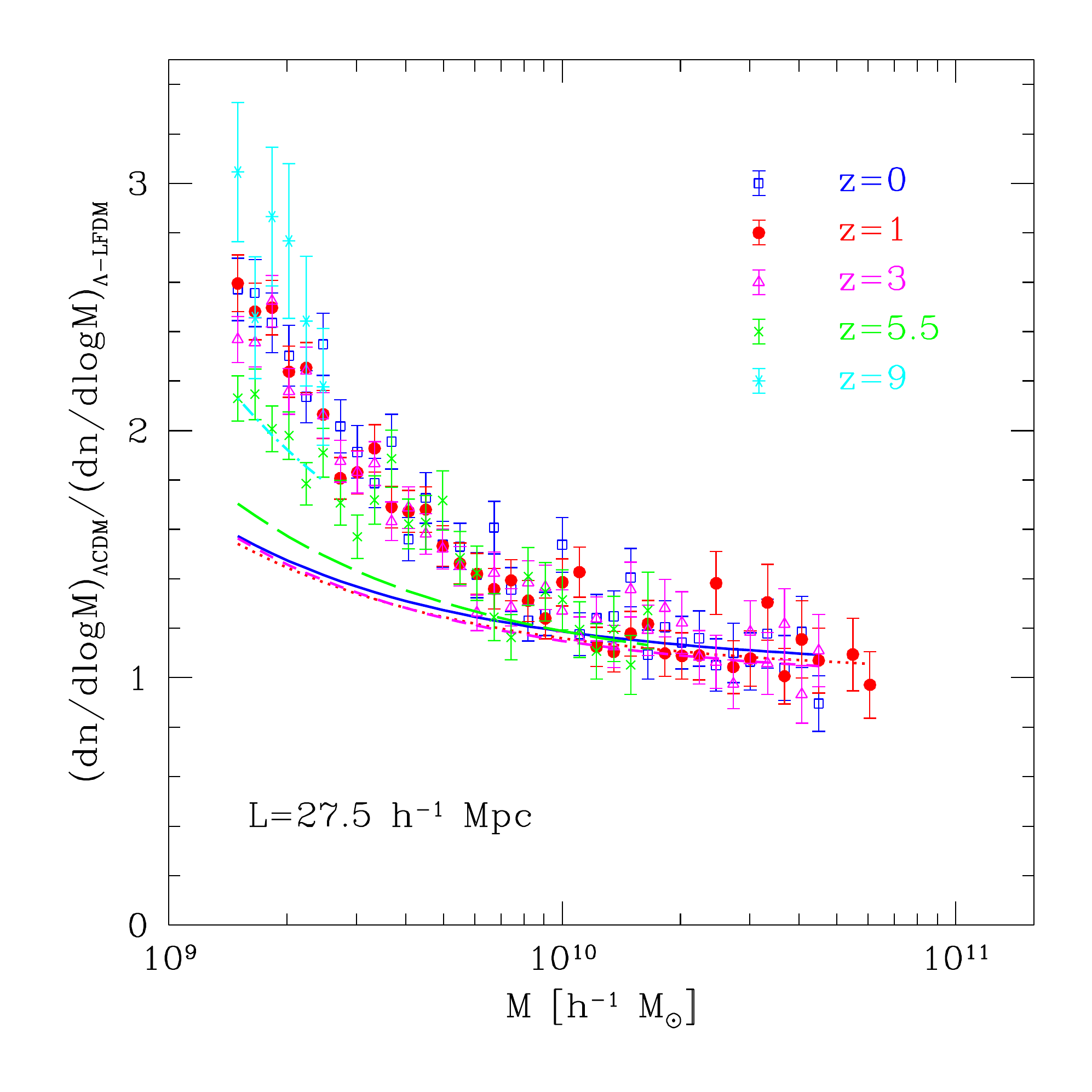}}
\caption{Left panel: Ratio of the multiplicity function of the $\Lambda$CDM model to that of the $\Lambda$-LFDM from the $(27.5\,{\rm h^{-1} Mpc})^3$ volume simulations with $512^3$ particles in mass bins of size $\Delta M/M=0.1$ containing at least $100$ halos such that finite volume effects are negligible; consequently, the mass range reduces for increasing redshifts. The different points correspond to redshifts from $z=0$ to $9$ as in Fig.~\ref{fig4}. Error bars are given by the propagation of Poisson errors. The different lines (solid, short dashed, dotted, long-dashed, dot-dashed) correspond to the ratio at different redshifts ($0$ to $9$, respectively) obtained assuming the Sheth-Tormen multiplicity function with parameters calibrated at $z=0$ for the $\Lambda$CDM and $\Lambda$-LFDM models. Right panel: Ratio of the $\Lambda$CDM mass function to that of the $\Lambda$-LFDM model in mass bins of size $\Delta M/M=0.1$ containing at least $100$ halos from the $(27.5\,{\rm h^{-1} Mpc})^3$ box simulations, as in the left panel.}\label{fig5}
\end{centering}
\end{figure*}

We can see that despite the suppressed amplitude of the linear matter power spectrum at small scales (see right panel in Fig.~\ref{fig1}), the halo formation proceeds as in a hierarchical bottom-up scenario, and the abundance of halos increases from high to low redshifts as a function of halo mass. At $z=0$, the ST mass function provides a good fit to the N-body measurements to better than $10\%$ in the mass range $10^{10}<M\,[\rm h^{-1} M_\odot]<2\times 10^{12}$. For larger masses, deviations from the ST fit are dominated by finite volume errors. In contrast, at lower masses, the ST overestimates the abundance of low-mass halos with deviations up to $30\%$. At redshifts $z=1,3$ and $5$, such deviations increase and extend over the entire range of masses with deviations as large as $50\%$ in the range where volume effects are negligible. At $z=9$, the low-mass end is still overestimated, while the abundance of halos with mass $M\gtrsim 10^{10}\,{\rm h^{-1} M_\odot}$ is underestimated. This trend is a manifestation of departure from the universality of the halo mass function already found in standard CDM cosmologies (see e.g. Ref.~\cite{Tinker2008,Crocce2010,Courtin2011}). Nevertheless, the fact that ST systematically overestimates the abundance of low-mass halos ($M\lesssim 10^{10}\,{\rm h^{-1} M_\odot}$) suggests a substantial departure of the nonlinear gravitational dynamics from that of the standard CDM ellipsoidal collapse model encoded in the ST multiplicity function (see the Appendix). To see this more clearly we compare the multiplicity function of the $\Lambda$-LFDM model to that of the $\Lambda$CDM. For this we use data from the $(27.5\,{\rm h^{-1} Mpc})^3$ volume simulation with $512^3$ particles. As with the case of the relative difference between the matter power spectra presented earlier, the ratio of the multiplicity functions is less affected by mass resolution errors.

In the left panel of Fig.~\ref{fig5}, we plot the ratio of the multiplicity functions at $z=0,1,3,5.5$ and $9$ in mass bins of size $\Delta M/M=0.1$ containing at least $100$ halos. This limits the mass range of interest to an interval where volume effects remain negligible, and thus it reduces from $z=0$ to $z=9$, as volume effects are largest at the high-mass end. We can see that independently of the redshift, for $M\gtrsim 10^{10}\,{\rm h^{-1} M_\odot}$ the multiplicity functions of the two simulated cosmologies are identical to numerical precision. We may also notice that in this mass range the ratio of the ST multiplicity functions calibrated to the data at $z=0$ for the two cosmological models reproduces fairly well the numerical results. This is not the case at lower masses, where ST predictions largely deviate from the N-body results. This suggests that the use of standard mass function formulas developed in the framework of the CDM paradigm to infer predictions for nonstandard CDM models characterized by a cutoff in the linear matter power spectrum at small scales may lead to significant misestimates of the halo abundances at low mass. 

Let us now turn to the comparison of the halo abundance between $\Lambda$CDM and $\Lambda$-LFDM. In the right panel of Fig.~\ref{fig5}, we plot the mass function of $\Lambda$CDM halos relative to that of the $\Lambda$-LFDM at $z=0,1,3,5.5$ and $9$ in mass bins of size $\Delta M/M=0.1$ containing at least $100$ halos. Again, the ratio is of order unity at all simulated redshifts for $M\gtrsim 10^{10}\,{\rm h^{-1} M_\odot}$ which is consistent with the fact that in this range the multiplicity functions of the two models are identical. Furthermore, in this range also the variance of the smoothed linear density field predicted by the two cosmologies is nearly identical, thus leading to a similar halo abundance well reproduced by the ST mass function. In contrast, we can see that for $M\lesssim 10^{10}\,{\rm h^{-1} M_\odot}$, halos in $\Lambda$CDM are up to a factor of $3$ more abundant than in $\Lambda$-LFDM. This is due to the differences in the multiplicity function shown in the left panel of Fig.~\ref{fig5} and amplified by the fact that on these scales the smoothed linear density field of the $\Lambda$-LFDM model is suppressed compared to the $\Lambda$CDM case. Also notice that there is no redshift evolution of the relative abundance between $z=3$ and $0$, while a modest evolution occurs at higher redshifts for halos with mass $M\sim 10^{9}\,{\rm h^{-1} M_\odot}$. This is consistent with the fact that in $\Lambda$-LFDM, small-mass halos form later than in $\Lambda$CDM, thus leading to differences that can be tested through high-redshift universe observations as already explored in the case of WDM models in Ref.~\cite{Schultz2014}. The lower abundance of low-mass halos in LFDM models compared to $\Lambda$CDM is a direct consequence of the suppression of power at small scale in the linear matter power spectrum beyond the cutoff scale. Since in the LFDM scenario the location of the cutoff depends on the epoch of dark matter formation, we can expect that DM forming earlier than that in the model considered here will decrease the discrepancy with respect to the $\Lambda$CDM model, while a later formation will increase it. Thus, measurements of the abundance of dwarf galaxies may provide direct constraints on the epoch of dark matter formation in LFDM models.

\section{Halo Density Profile}\label{profiles}
\begin{figure}
\includegraphics[scale=0.4]{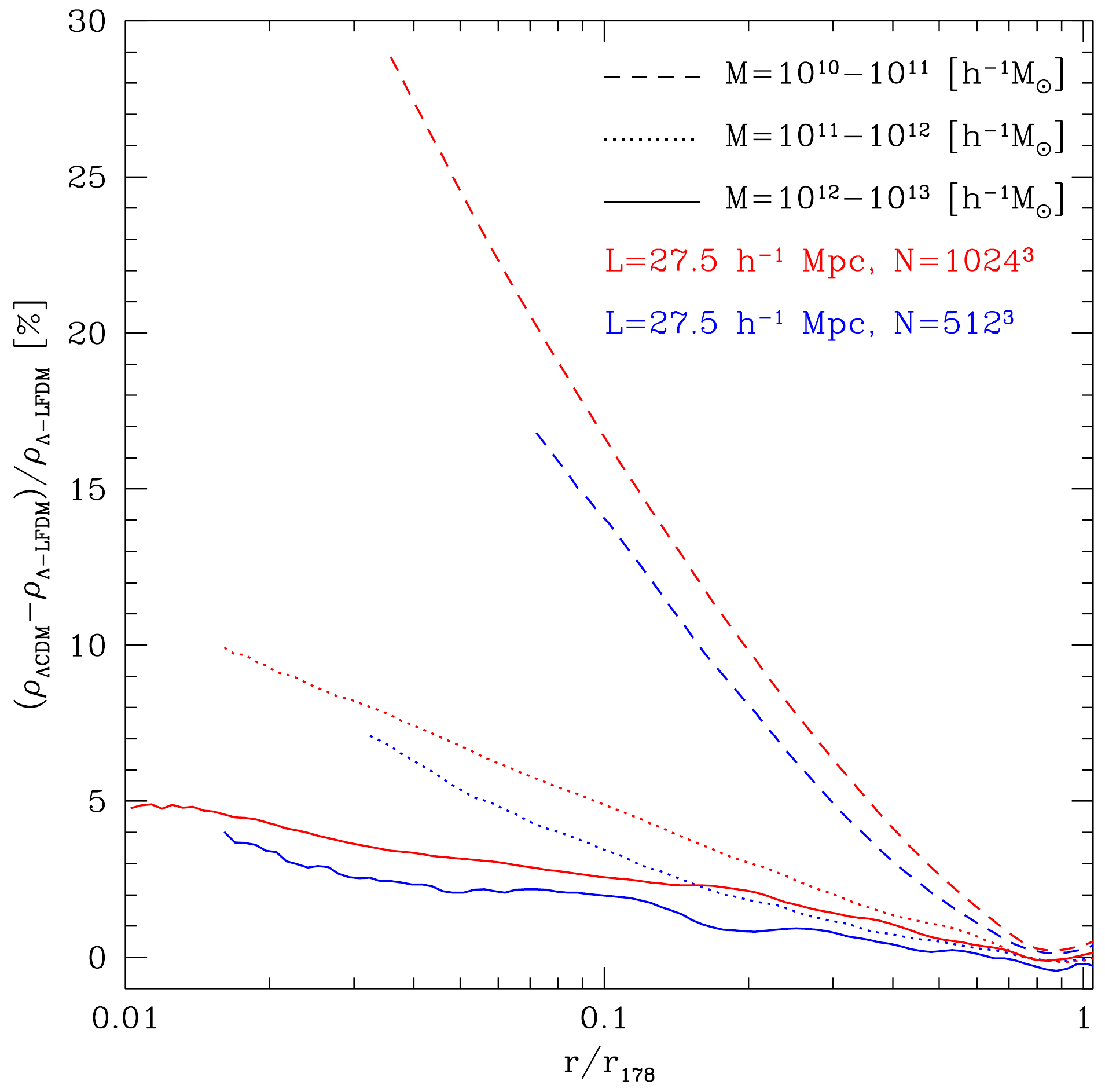}
\caption{Relative difference of the averaged halo density profile at $z=0$ in $\Lambda$CDM with respect to the $\Lambda$-LFDM model, as a function of radius normalized by $r_{178}$. The averaged profiles are obtained by stacking the halo profiles within mass bins $M=10^{10}-10^{11}$ h$^{-1}$ M$_\odot$ (dashed line), $10^{11}-10^{12}$ h$^{-1}$ M$_\odot$ (dotted line) and $10^{12}-10^{13}$ h$^{-1}$ M$_\odot$ (solid line) for the $(27.5\,{\rm h^{-1} Mpc})^3$ box simulations with $512^3$ (blue lines) and $1024^3$ (red lines) particles.}\label{fig6}
\end{figure}

We now focus on the density profile of DM halos. In Fig.~\ref{fig6} we plot the relative difference of the $\Lambda$CDM averaged halo density profile at $z=0$ with respect to that of the $\Lambda$-LFDM case for halos in three different mass bins. Within each mass bin, individual halo profiles are stacked to obtain the averaged profile. To facilitate stacking, radius is normalized by $r_{178}$ corresponding to an enclosed overdensity $\Delta=178$ relative to the cosmic mean matter density. To control numerical resolution effects, we plot results from the $(27.5\,{\rm h^{-1} Mpc})^3$ box simulations with $512^3$ (blue lines) and $1024^3$ (red lines) particles. Moreover, we limit the curves up to the physical scale associated with the size of the most refined cell of the simulations. We can see that the lower-resolution simulations systematically underestimate the profile differences with respect to the higher-resolution runs with deviations up to $\sim 3\%$ level.

We find that density profiles for halos with mass $M>10^{11}\,{\rm h^{-1} M_\odot}$ are nearly identical in the two cosmologies for $r/r_{178}\gtrsim 0.1$. At smaller radii, deviations do not exceed the $10\%$, level with $\Lambda$CDM halos being on average denser than in $\Lambda$-LFDM. In the case of halos with mass in the range $M=10^{10}-10^{11}\,{\rm h^{-1} M_\odot}$, we find larger deviations. In particular, in the CDM case the averaged profile at radii $r/r_{178}\lesssim 0.1$ is $\sim 20\%$ denser than the LFDM counterpart. As already noticed in Ref.~\cite{Buckley2014}, this is consistent with the fact that the two models only differ for the amplitude of the initial power spectrum at small scales. In fact, as shown by the analysis of the mass function, the suppression of power at small scales in the $\Lambda$-LFDM model causes small-mass halos to form later than in $\Lambda$CDM, thus resulting in lower density profiles. Halos of higher mass, on the other hand, assemble nearly at the same time in the two cosmologies, thus leading to nearly identical profiles. If the trend shown in Fig.~\ref{fig6} at small radii is extrapolated to lower-mass halos, then these results do not exclude the possibility that halos with mass $M<10^{10}\,{\rm h^{-1} M_\odot}$ may have cored profiles; however, we are not able to address this point at the moment, since our simulations do not possess the required resolution.

\section{Circular Velocity Distribution}\label{mcvf}
The abundance of dark matter halos can be probed through measurements of the galaxy luminosity function provided prior knowledge of the relation between galaxy luminosity and halo mass exists. However, this is hard to predict, since it depends on the galaxy formation process itself. In contrast, the velocity function, i.e. the number density of galaxies with a given circular velocity, can be predicted more reliably and tested against observations. It is indeed the comparison of the circular velocity of the satellites of the Local Group against N-body simulation results that originally pointed to the missing satellite problem \cite{Klypin1999,Moore1999}. Another advantage of the velocity function is that it probes the abundance of all virialized structures. In the past few years, analyses of the velocity function inferred from a number of surveys of the local volume have pointed to suppressed abundances of low-velocity field galaxies compared to expectations from $\Lambda$CDM simulations \cite{Zavala2009,Papastergis2011,Papastergis2015}. These findings have been recently confirmed by the measurements of the circular velocity distribution in galaxies within $10$ Mpc from the Local Volume catalog \cite{Klypin2014}. The circular velocity distribution is indeed a promising tool to constrain DM properties; however, it is worth mentioning that any analysis using DM-only simulations can introduce a mass-dependent bias, since the measured circular velocity is that of the baryon component inside dark matter halos.

Here, we compare the measurements of the circular velocity distribution from Ref.~\cite{Klypin2014} against estimates from our $\Lambda$CDM and LFDM halo catalogs of the $(27.5\,{\rm h^{-1} Mpc})^3$ volume simulations with $1024^3$ particles. To derive predictions from DM-only simulations that can be compared to observations, we follow the procedure described in Ref.~\cite{Klypin2014}. More specifically, we correct the estimated halo circular velocities to account for the effect of baryons using the prescription described in Ref.~\cite{Klypin2014} and based on the analysis by Ref.~\cite{TrujilloGomez2011}. Then, we multiply the estimated velocity functions by a factor $1.25$ to account for the contribution of subhalos not included in our numerical catalogs. As pointed out in Ref.~\cite{Klypin2014}, this factor corresponds to the fraction of subhalos in $\Lambda$CDM for velocities $V\lesssim 200$ km s$^{-1}$. In the LFDM scenario, the fraction of subhalos may be even smaller due to the power suppression at small scales; thus the choice adopted here is a rather conservative one.

\begin{figure}
\includegraphics[scale=0.4]{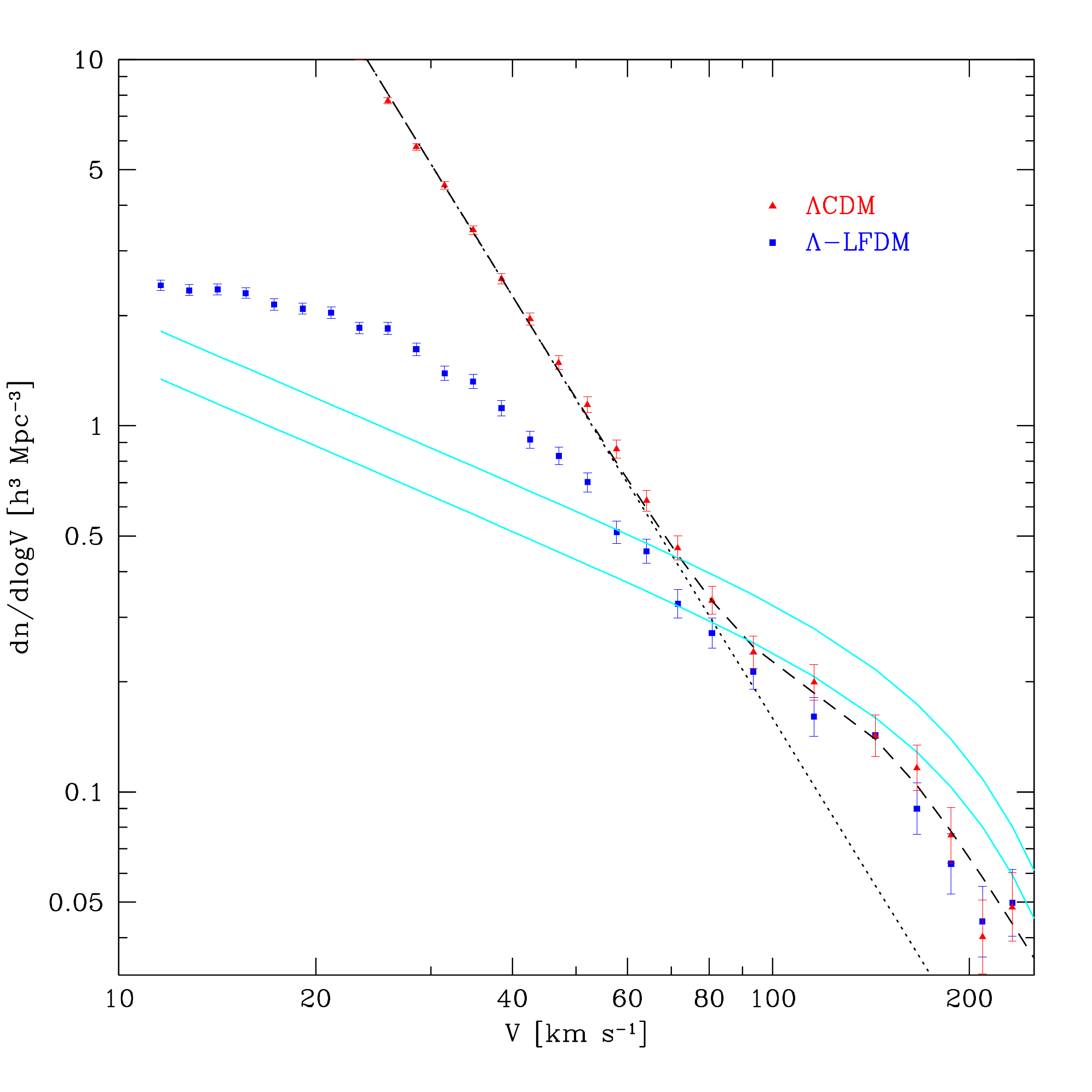}
\caption{Differential circular velocity function for the $(27.5\,{\rm h^{-1} Mpc})^3$ box simulations with $1024^3$ particles for the $\Lambda$CDM (red triangles) and $\Lambda$-LFDM (blue squares) models. Error bars are given by Poisson errors. The dotted straight line is the fitting function to the $\Lambda$CDM simulation given by Eq.~(5) in Ref.~\cite{Klypin2014}, while the dashed line includes the baryonic corrections. The cyan solid lines denote the region of the observed velocity function to within $15\%$ error from Ref.~\cite{Klypin2014}.}\label{fig7}
\end{figure}

In Fig.~\ref{fig7}, we plot the differential circular velocity function $dN/d\log{V}$ for the $\Lambda$CDM (red triangles) and $\Lambda$-LFDM (blue squares) models. The dotted straight line is the fitting function to the $\Lambda$CDM simulation given by Eq.~(5) in Ref.~\cite{Klypin2014}, while the dashed line includes the baryonic corrections. Since the cosmological parameters considered here are marginally different from Ref.~\cite{Klypin2014}, we normalize these fitting functions to the circular velocity function obtained from our $\Lambda$CDM simulation. The cyan solid lines enclose the region of the observed velocity function (with a fitting function given by Eq.~(12) in Ref.~\cite{Klypin2014}) to within $15\%$ error. As we can see, the LFDM model predicts a lower abundance of low-velocity galaxies in better agreement with observations than $\Lambda$CDM. Furthermore, since the suppression is related to the transition redshift $z_t$ of DM formation (the later the transition, the larger the suppression of low-mass halo abundances), it is possible that setting $z_t$ to a value slightly smaller than that assumed here may reproduce the observed velocity distribution up to statistical errors.

\section{Conclusions}\label{conclu}
The distribution of matter at small scales may provide insights on the nature of dark matter particles and potentially constrain scenarios alternative to the standard cold dark matter paradigm. In this work we have performed a first study of the nonlinear clustering of late-forming dark matter using a suite of high-resolution N-body simulations. The main phenomenological feature of this scenario is the presence of damped acoustic oscillations at large wave numbers in the linear matter power spectrum. The scale where such a power suppression occurs depends on the epoch of the phase transition of a scalar field coupled to relativistic particles that generates the DM particles. Without requiring an extreme fine-tuning of the microscopic model parameters, this may occur before matter-radiation equality and imprint a pattern of damped oscillations at $k\gtrsim 10$ h Mpc$^{-1}$ in the initial conditions. Because of this, LFDM models are indistinguishable from standard CDM on large scales, while they are constrained by probes of the small-scale clustering of matter.

From the analysis of N-body simulations, we have shown that LFDM could be a viable alternative to CDM, as it provides predictions of the nonlinear matter power spectrum largely consistent with bounds from high-redshift Lyman-$\alpha$ power spectrum measurements. It should be noted that the Lyman-$\alpha$ constraints have so far been obtained for a restricted class of DM models (such as WDM) with a sharp cutoff in the linear power spectrum. As LFDM models have spectra with considerably shallower slopes than WDM, it is very likely that LFDM models that resolve $\Lambda$CDM inconsistencies on galactic and subgalactic scales would also be in better agreement with Lyman-$\alpha$ measurements.

We have evaluated the halo abundance; comparison with predictions from the standard $\Lambda$CDM model shows that small-mass LFDM halos ($10^9<M\,[{\rm h^{-1} M_\odot}]<10^{10}$) are less abundant than their CDM counterparts. We have also evaluated the corresponding differences in the circular velocity function at $z=0$ and shown that the LFDM scenario is in better agreement with observations of the number density of low-velocity galaxies than $\Lambda$CDM. The number density of low-mass halos varies with redshift, with low-mass halo abundance up to $\sim 3$ smaller than $\Lambda$CDM at $z>3$. This indicates that halos in this mass range form later than in $\Lambda$CDM, thus leading to lower densities in the internal part of the halos as confirmed by the study of density profiles. On the other hand, the assembly of more massive halos occurs as in $\Lambda$CDM, and we find no significant differences in their density profiles between the two cosmologies.

These results confirm previous numerical studies of nonstandard DM models characterized by initial linear matter power spectra similar to that of LFDM. Since LFDM behaves as a collisionless component, the lack of cored profiles in Milky Way--like halos seems inherently related to the fact that in order to satisfy the Lyman-$\alpha$ constraints, the suppression of power in the initial power spectrum must occur on wave number $k>10$ h Mpc$^{-1}$. This suggests that only by including DM self-interactions would it be possible to develop cored profiles. On the other hand, one should note that baryonic feedback does play a role at these scales. Given the differences in the abundance and assembly of small-mass halos in $\Lambda$-LFDM and $\Lambda$CDM, it will be of interest to investigate the dynamics of baryons in the LFDM scenario. This may leave distinct observational features in the formation of stars and galaxies which warrant further investigation.

\begin{acknowledgments}
We would like to thank Matteo Viel and Jes\'{u}s Zavala for useful comments and suggestions. This work was granted access to HPC resources of IDRIS through time allocations made by GENCI (Grand \'Equipement National de Calcul Intensif) on the machine ADA No. x2014042287. We acknowledge support from the DIM ACAV of the Region \^Ile-de-France. The research leading to these results has received funding from the European Research Council under the European Community's Seventh Framework Programme (FP7/2007-2013 Grant Agreement No. 279954).
\end{acknowledgments}

\appendix*
\section{MULTIPLICITY FUNCTION}\label{mult}
In the Press-Schechter approach \cite{PS} the halo mass function can be written as
\begin{equation}
\frac{dn}{dM}=\frac{\rho_m}{M}\frac{d\,\textrm{ln}\sigma^{-1}}{dM}f(\sigma),
\end{equation}
where $\rho_m$ is the mean cosmic matter density and $\sigma$ is the root-mean-square fluctuation of the linear matter density field smoothed on a scale $R(M)$ enclosing a mass $M$ with
\begin{equation}
\sigma^2(M)=\frac{1}{2\pi^2}\int dk k^2 P(k)W^2[k,R(M)],
\end{equation}
where $P(k)$ is the linear DM power spectrum and $W(k,R)$ is the Fourier transform of the smoothing function in real space (here we consider a top hat filter). The function $f(\sigma)$ is the so-called ``multiplicity function'' which carries all information on the nonlinear gravitational processes that lead to the formation of halos. 

In the framework of the excursion set theory \cite{Bond1991}, this can be computed from the distribution of random walks first crossing a collapse density threshold that encodes the nonlinear gravitational dynamics of matter collapse. 

It was shown in Ref.~\cite{ShethMoTormen2001} that the multiplicity function for uncorrelated walks with a threshold barrier motivated by the ellipsoidal collapse model is well approximated by a the Sheth-Tormen function (apart from a fudge factor coefficient necessary to recover agreement with the N-body results) given by \cite{ST1999}
\begin{equation}
f_{ST}(\sigma)=A\sqrt{\frac{2a}{\pi}}\frac{\delta_c}{\sigma}\left[1+\left(\frac{\delta_c\sqrt{a}}{\sigma}\right)^{-2p}\right]e^{-a\delta_c^2/2\sigma^2},
\end{equation}
where $\delta_c$ is the linearly extrapolated spherical collapse density threshold. Since after the transition LFDM behaves as a collisionless component, the spherical collapse dynamics is identical to that of CDM\footnote{This might not be the case for the ellipsoidal collapse model which explicitly depends on the mass of the collapsing object, while the spherical collapse is independent of the mass.}. At $z=0$ we have $\delta_c=1.673$, and we compute the exact redshift dependence $\delta_c(z)$ by numerically solving the spherical collapse equations as in Ref.~\cite{Courtin2011}.

We calibrate the ST parameters $A$, $a$ and $p$ to the numerical mass function at $z=0$ from the $(27.5\,{\rm h^{-1} Mpc})^3$ volume simulation of the $\Lambda$CDM and $\Lambda$-LFDM model. We find $A=0.145$ and $a=0.695$ for both models, while $p=0.15$ for $\Lambda$CDM and $p=0.1$ for $\Lambda$-LFDM. As pointed out in Ref.~\cite{ShethMoTormen2001}, the parameter $p$ is determined by the shape of the mass function at the low-mass end, which in turn depends on the form of the ellipsoidal collapse threshold (see also Refs.~\cite{CA2011,Lapi2013} for an explicit relation between the mass function and the parameters of an ellipsoidal-collapse-inspired barrier). Hence, the large deviations from unity of the ratio of the multiplicity functions at low masses shown in the left panel of Fig.~\ref{fig5} are clearly indicative of a departure of the nonlinear collapse dynamics of the $\Lambda$-LFDM model from that of the $\Lambda$CDM.

\end{document}